\documentclass[
 reprint,
superscriptaddress,
amsmath,amssymb,
aps,
pra,
]{revtex4-2}

\usepackage{graphicx}
\usepackage{dcolumn}
\usepackage{bm}
\usepackage{url}
\usepackage{multirow}
\usepackage{lipsum}
\usepackage{hyperref}
\usepackage[utf8]{inputenc}		
\usepackage{lmodern} 
\usepackage{xcolor}
\usepackage{microtype}
\usepackage{caption}
\usepackage{verbatim}
\usepackage{subcaption}
\usepackage[percent]{overpic}

\usepackage{xcolor}

\begin{document}

\preprint{APS/123-QED}

\title{Addressing the impact of surface roughness on epsilon-near-zero substrates}

\author{David Navajas}
\affiliation{%
 Department of Electrical, Electronic and Communications Engineering, Institute of Smart Cities (ISC), Public University of Navarre (UPNA), 31006 Pamplona, Spain
}
\author{José M. Pérez-Escudero}
\affiliation{%
 Department of Electrical, Electronic and Communications Engineering, Institute of Smart Cities (ISC), Public University of Navarre (UPNA), 31006 Pamplona, Spain
}
\author{María Elena Martínez-Hernández
}
\affiliation{%
 Department of Electrical, Electronic and Communications Engineering, Institute of Smart Cities (ISC), Public University of Navarre (UPNA), 31006 Pamplona, Spain
}
\author{Javier Goicoechea}
\affiliation{%
 Department of Electrical, Electronic and Communications Engineering, Institute of Smart Cities (ISC), Public University of Navarre (UPNA), 31006 Pamplona, Spain
}
\author{Iñigo Liberal}
\thanks{Corresponding author: inigo.liberal@unavarra.es}%
\affiliation{%
 Department of Electrical, Electronic and Communications Engineering, Institute of Smart Cities (ISC), Public University of Navarre (UPNA), 31006 Pamplona, Spain
}%

\begin{abstract}
Epsilon-near-zero (ENZ) media have been very actively investigated due to their unconventional wave phenomena and strengthened nonlinear response. However, the technological impact of ENZ media will be determined by the quality of realistic ENZ materials, including material loss and surface roughness. Here, we provide a comprehensive experimental study of the impact of surface roughness on ENZ substrates. Using silicon carbide (SiC) substrates with artificially induced roughness, we analyze samples whose roughness ranges from a few to hundreds of nanometer size-scales. It is concluded that ENZ substrates with roughness in the few nanometer scale are negatively affected by coupling to longitudinal phonons and strong ENZ fields normal to the surface. On the other hand, when the roughness is in the hundreds of nanometer scale, the ENZ band is found to be more robust than dielectric and surface phonon polariton (SPhP) bands. 
\end{abstract}

\maketitle

\section{Introduction}

Epsilon-near-zero (ENZ) media, i.e., materials and/or metamaterial constructs with a near-zero permittivity, have become a very active field of research field due to their qualitatively different optical behaviour \cite{Liberal2017rev,Kinsey2019rev,Reshef2019nonlinear,Ziolkowski2004propagation}. A variety of wave phenomena emerges from the different physics of ENZ media, examples including supercoupling \cite{Silveirinha2006,Edwards2008experimental} and ideal fluid flow \cite{Liberal2020ideal,Li2022ideal}, geometry-invariant resonators \cite{Liberal2016geometry}, directive emission \cite{Enoch2002,Alu2007directive}, photonic doping \cite{Liberal2017photonic}, nonradiating modes \cite{Silveirinha2014trapping,Monticone2014embedded,Liberal2016nonradiating} and guided modes with a flat dispersion profile \cite{Vassant2012berreman,Campione2015theory,Runnerstrom2017epsilon}, just to name a few. Moreover, ENZ media intrinsically enhances light-matter interactions, as it is the case for nonlinear optics \cite{Alam2016large,Caspani2016enhanced,Capretti2015enhanced,Khurgin2021,Yang2019}, electrical \cite{Huang2016gate,Wood2018gigahertz} and optical \cite{Kinsey2015epsilon,Bohn2021all} modulation, spontaneous emission \cite{Lobet2020,So2020,Li2016controlling,Fleury2013}, magnon-optical photon coupling \cite{Bittencourt2022optomagnonics,Bittencourt2022light}, entanglement generation \cite{Li2021multiqubit,Ozguun2016tunable}, and light concentration on ultra-thin metallic films for thermal emitters \cite{Navajas2022,Perez2020} and optoelectronic \cite{Krayer2019optoelectronic} devices. 

Due to the scientific and technological interest of ENZ media, there has been an intensive research on material platforms exhibiting a near-zero permittivity at infrared, visible and even ultraviolet frequencies. Similar to plasmonic systems, the performance of ENZ materials at optical frequencies is limited by material loss \cite{Khurgin2015deal,West2010searching}. A detailed account of materials with an ENZ response, their frequency of operation, and their ranking in accordance to material losses can be found in popular reviews on the topic \cite{Kinsey2019rev}.

\begin{figure}
    \centering
    \includegraphics[width=\linewidth]{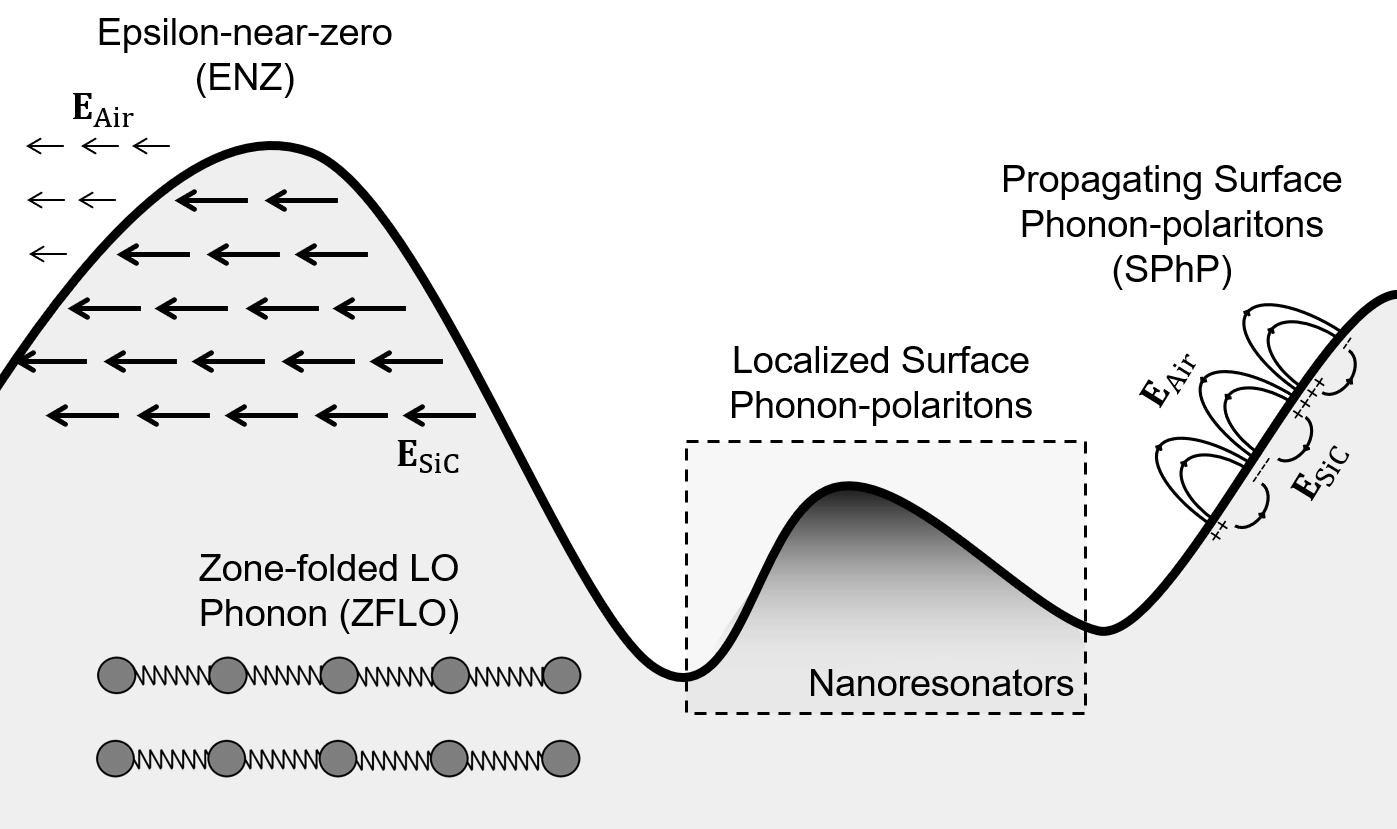}
    \caption{{\bf Polaritonic phenomena excited at a silicon carbide (SiC) substrate with artificially-induced roughness.} Strong longitudinal epsilon-near-zero (ENZ) fields, propagating and localized surface phonon-polaritons (SPhP), and coupling to zone-folded longitudinal phonons (ZFLO).}
    \label{Fig:ConceptIntro}
\end{figure}

Beyond material losses, surface roughness is one of the major challenges in the development of high-performance nanophotonic technologies. For example, surface roughness severely affects plasmonic systems \cite{Nagpal2009ultrasmooth,Malureanu2015ultra}, as they are based on surface modes tightly confined to a metal interface. Sidewall roughness is also the major loss mechanism in state-of-the-art integrated photonic resonators and waveguides \cite{Ji2017ultra,Ji2021exploiting}. However, the impact of surface roughness on ENZ media is far less studied. 
From the known theory of ENZ media, two apparently contradictory arguments on the impact of surface roughness on ENZ media are possible. On the one hand, due to the continuity of the normal displacement field, $\mathbf{D}=\varepsilon \mathbf{E}$, oblique incidence over an ENZ interface results in the excitation of strong longitudinal electric fields (see Fig.\,\ref{Fig:ConceptIntro}). In fact, longitudinal modes would be supported in an ideal ENZ medium \cite{Landau2013}. Thus, it has been suggested that the response of ENZ substrates would be particularly affected by surface roughness \cite{Javani2016real}, due to the excitation of strong longitudinal fields at a rough boundary. On the other hand, ENZ media is characterized by an effective enlargement of the wavelength, $\lambda_{\rm eff}=\lambda_0/\sqrt{\varepsilon}\gg 1$. Therefore, it might be expected that variations of the geometry would only lead to small changes in the response of the system, in line with geometry-invariant phenomena observed in ENZ media \cite{Liberal2016zero}.

In this work, we address the impact of surface roughness on ENZ substrates by experimentally and numerically investigating the reflectivity of a silicon carbide (SiC) substrate with different levels of artificially-induced roughness. Silicon carbide is a polar dielectric, whose response at infrared frequencies is characterized by a reflective band arising from the coupling to optical phonons  \cite{Caldwell2015}. The excellent optical properties of SiC have facilitated experimental demonstrations of superlensing \cite{Taubner2006near}, extraordinary transmission \cite{Urzhumov2007}, thermal emitters driven by waste heat \cite{Lu2020narrowband} and parametric amplification of phonons \cite{Cartella2018}.

SiC is also an excellent material platform for investigating the impact of surface roughness on ENZ media for two reasons: First, it has a high-quality ENZ response, which has enabled the experimental demonstration of frequency pinning of resonant nanoantennas \cite{Kim2016role}, ENZ high-impedance thermal emitters \cite{Perez2020,Navajas2022}, and ENZ waveguide and cavity modes \cite{Folland2020vibrational,Yoo2021ultrastrong}. Second, its permittivity has a Lorentzian dispersion profile, including a band of negative permittivity supporting the propagation of surface phonon polaritons (SPhP), as well as frequency bands with a dielectric response. As schematically depicted in Fig.\ref{Fig:ConceptIntro}, a variety of polaritonic phenomena is excited at a rough SiC substrate, including distinct {\bf ENZ field} distributions, {\bf propagating} and {\bf localized SPhP}, and coupling to {\bf zone-folded longitudinal phonons (ZFLO)}. Consequently, SiC enables a direct comparison of the impact of surface roughness on ENZ media, plasmonic-like systems and dielectric media, all within the same sample.

\section{Experimental results}

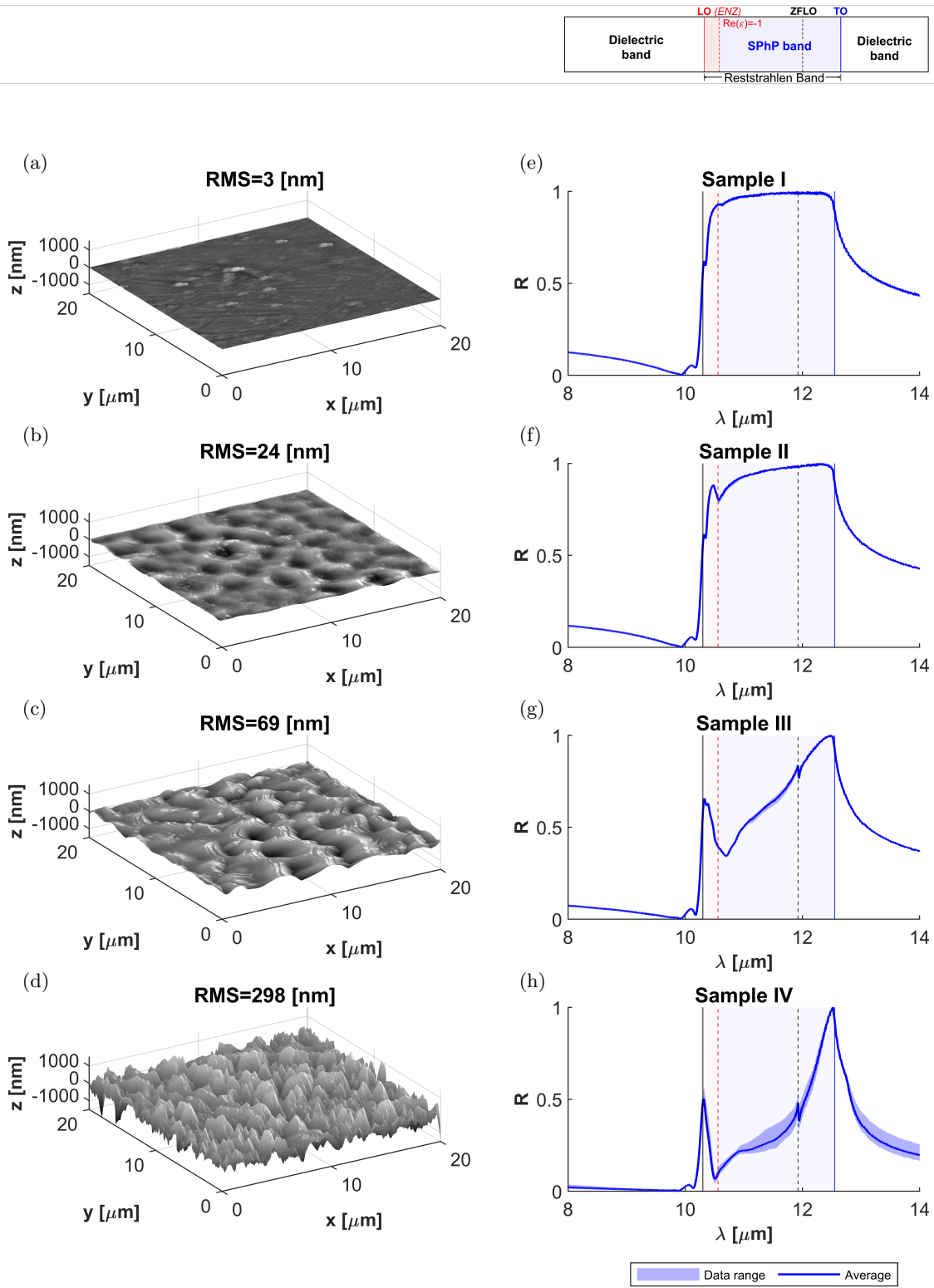
\begin{figure*}
\centering
    \begin{subfigure}{0.97\linewidth}
     \centering
    \includegraphics[width=\linewidth]{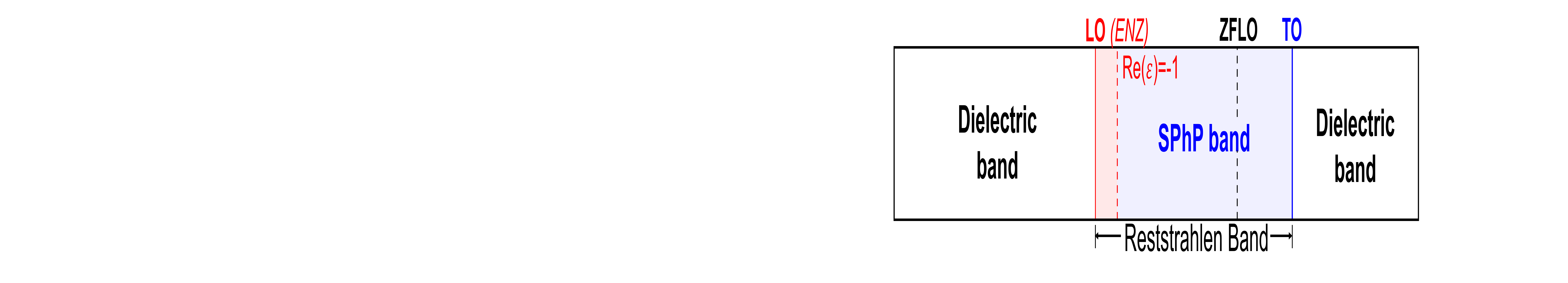}
    \end{subfigure}
    \begin{subfigure}{0.97\linewidth}
    \begin{overpic}[width=\linewidth]{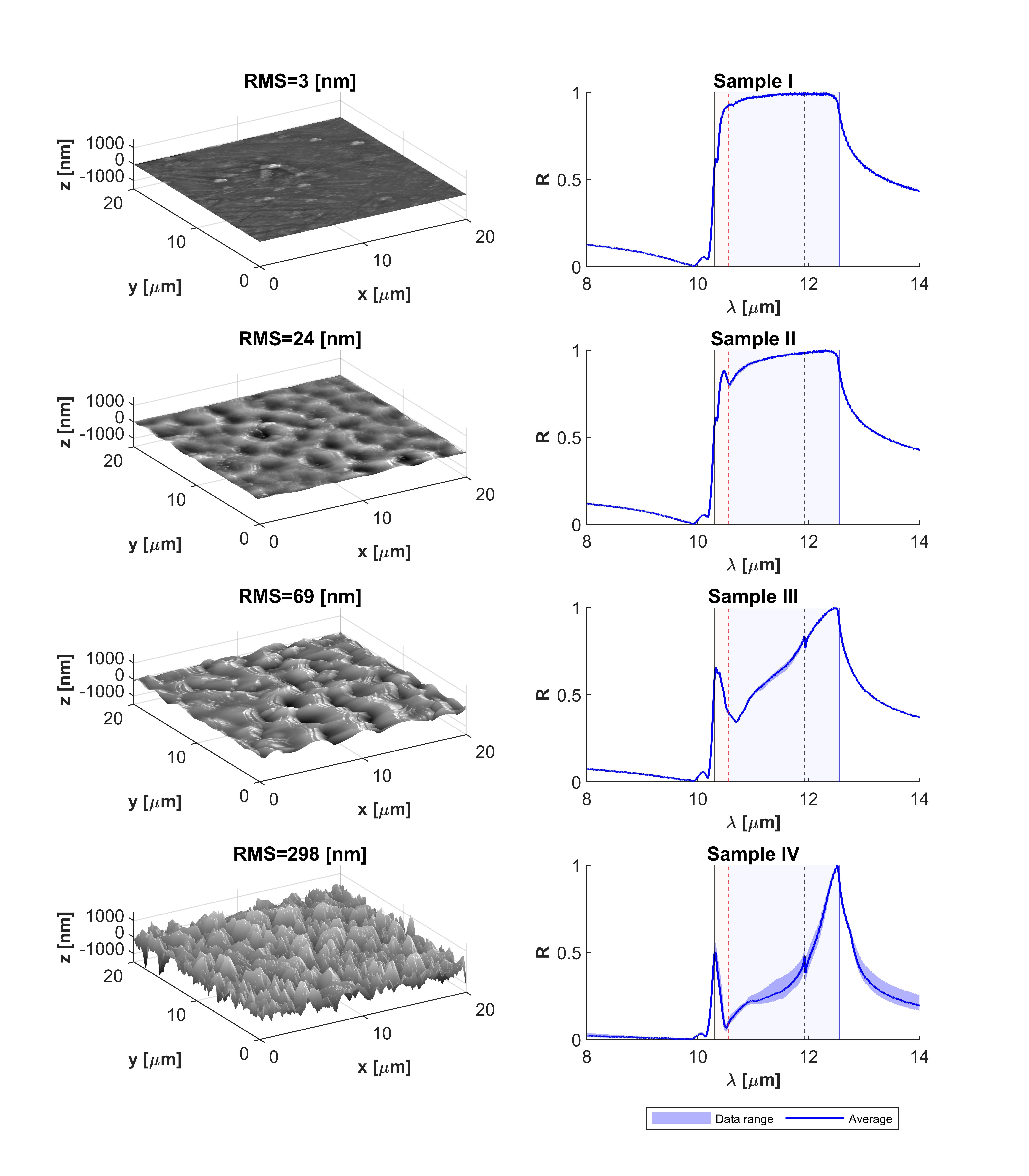}
    \put(6, 28){\small{(d)}}
    \put(6, 50){\small{(c)}}
    \put(6, 72){\small{(b)}}
    \put(6, 94){\small{(a)}}
    \put(46, 28){\small{(h)}}
    \put(46, 50){\small{(g)}}
    \put(46, 72){\small{(f)}}
    \put(46, 94){\small{(e)}}
    \end{overpic}
    \end{subfigure}
 \caption{{\bf Characterization of silicon carbide (SiC) substrates with artificially-induced roughness.} {\bf (a)-(d)} 3D Maps of the surface roughness obtained via atomic force microscopy (AFM).{\bf (e)-(h)} Reflectivity spectra obtained via FTIR spectroscopy. Each figure represents the average reflectivity obtained after four measurements on each sample. The data range expanded by all measurements is shown as a blue band between the minimum and maximum reflectivity measurements.}
\label{Fig:measurementssurfaces}
\end{figure*}

Artificial roughness was created on 220\,$\mu$m thick 4H-SiC wafers {\it via}  deep reactive ion etching (DRIE). Instead of optimizing the fabrication procedure to minimize roughness and sidewall angle in SiC substrates \cite{Luna2017deep,Luna2017dry,Racka2021review}, gas composition, time of exposure, pressure and temperature of the chamber were tuned to increase the root-mean-square (RMS) roughness. Specifically, the selected parameters were: No Helium in the chamber, 80 sccm of $SF_6$ and 10 sccm of $O_2$, 15 mTorr of pressure, temperature of 30\textdegree\,C and a time of exposure of 12 min (See Methods). We found that the ion bombardment results in the stochastic generation of significant roughness with large inter-sample variability, but low intra-sample variability. Atomic force microscopy (AFM) maps of the samples are shown in Figs.\,\ref{Fig:measurementssurfaces}(a)-(d), providing a more detailed view of the morphology of the samples. The root-mean-square (RMS) measurements of the samples are 3.0\,nm, 24.0\,nm, 68.8\,nm and 298.3\,nm. As we will show, this range of RMSs allows for the investigation of several regimes in the optical response of rough SiC substrates at infrared frequencies.

Reflectivity measurements were carried out {\it via}  Fourier transform infrared (FTIR) spectroscopy (see Methods), with the results gathered together in Figs.\,\ref{Fig:measurementssurfaces}(e)-(h). The reflectivity spectra provide information about the optical infrared response of the samples, including the coupling to different polaritonic modes, such as propagating and localized surface phonon polaritons (SPhP), as well as longitudinal (LO) and zone-folded (ZFLO) optical phonons. Figs.\,\ref{Fig:measurementssurfaces} (e)-(h) reports the average value and data range of the FTIR reflectivity for 4 different measurements on each sample. Individual data for each measurement is available in Supplementary Fig. 1. While it is found that the variability in the samples increases along with the surface roughness, the reflectivity measurements confirm a low intra-sample variability.

\begin{figure}
\centering
\begin{overpic}[width=0.95\linewidth]{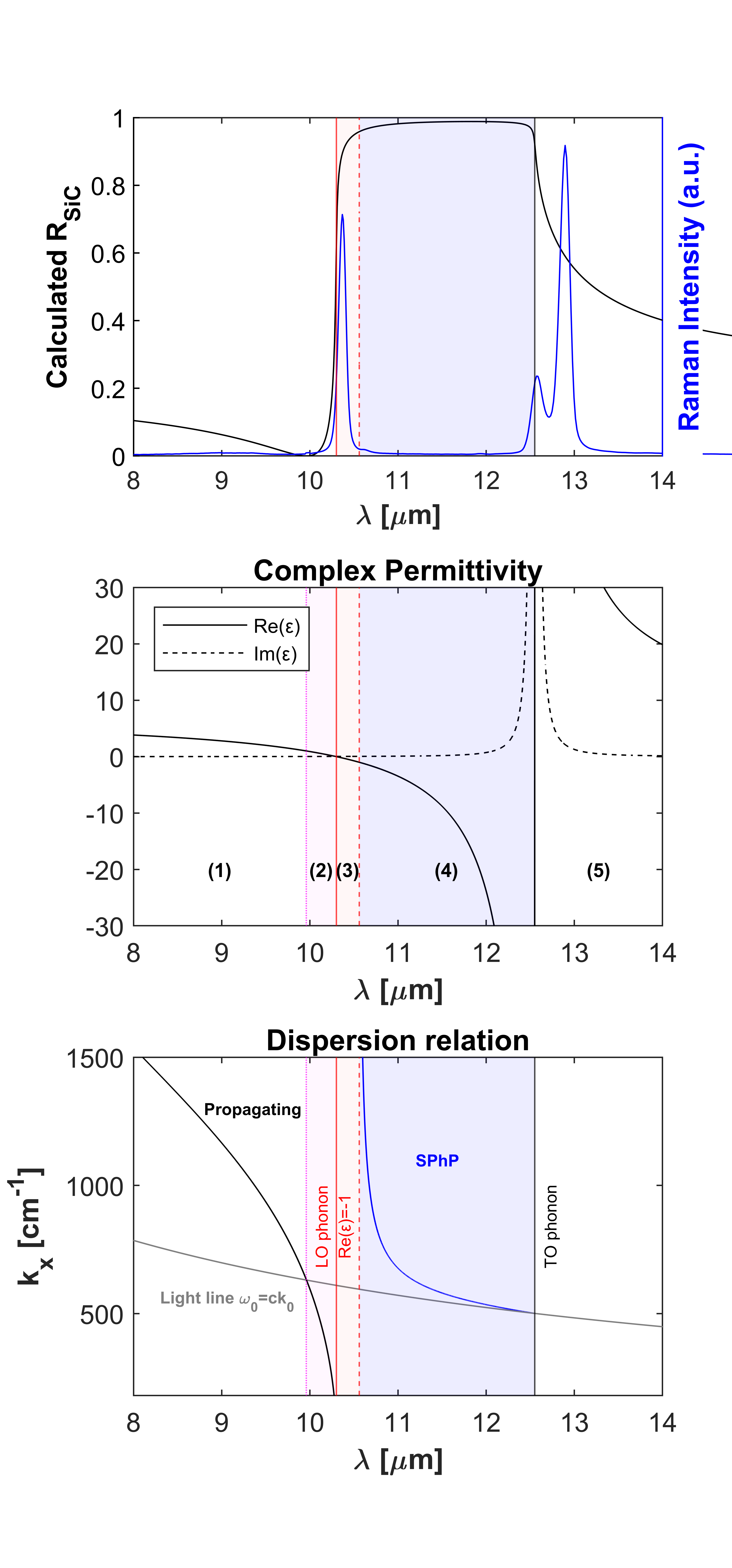}
    \put(6, 35){\small{(c)}}
    \put(6, 65){\small{(b)}}
    \put(6, 95){\small{(a)}}
    \end{overpic}
    \caption{{\bf Silicon carbide (SiC) theoretical dispersion properties.} (a) Theoretical reflectivity from a perfectly flat silicon carbide (SiC) substrate, and measurement of the Raman spectrum exhibiting phonon lines on both sides of the high-reflectivity band. (b) Real and imaginary parts of the SiC permittivity following a Lorentzian model. (c) Surface phonon polariton (SPhP) dispersion in the case of a smooth surface of SiC. The red highlighted zone represents the part where the phonon dispersion modes are still propagating. At $\varepsilon_{SiC}$=-1 (blue zone) the surface waves starts to be evanescent, and the SPhP phenomena starts.}
\label{Fig:evanescentpropagating}
\end{figure}

The measurements of {\bf Sample I}, the commercial wafer without etching featuring a RMS of 3\,nm, are presented in Fig.\,\ref{Fig:measurementssurfaces}(e). Overall, the reflectivity spectrum follows the theoretical response of a perfectly flat substrate. Specifically, it is characterized by a  highly reflective band, the Reststrahlen band, taking place between the transversal (TO, $\lambda_{\rm TO}=12.55\,\mu m$) and longitudinal (LO, $\lambda_{\rm LO}=10.3\,\mu m$) optical phonon wavelengths (see Fig.\,\ref{Fig:evanescentpropagating}(a)). As shown in Fig.\,\ref{Fig:evanescentpropagating}(b), the permittivity of SiC approximately follows a Lorentzian dispersion profile, and the Reststrahlen band corresponds to the frequency range where the real part of the permittivity of SiC is negative. The reflectivity also has a minimum at a wavelength of $\lambda=9.96\,\mu m$, where the real part of the permittivity approximately equals 1. 

Major deviations between the theory for a flat substrate and the reflectivity measurements for Sample I are the existence of three resonant dips at $\lambda_{\rm I}=10.63\,\mu m$, $\lambda_{\rm II}=10.35\,\mu m$ and $\lambda_{\rm III}=10.17\,\mu m$ wavelengths. 
The first resonant wavelength, $\lambda_{\rm I}=10.63\,\mu m$ lies near the wavelength the real part of the permittivity of SiC is expected to equal -1. Thus, it aligns the SPhP resonance, the edge where the band of propagating SPhP modes starts, and the wavelength with a larger density of propagating SPhP modes. For reference, the theoretical dispersion of SPhPs is reported in Fig.\,\ref{Fig:evanescentpropagating}(c). Therefore, this dip can be ascribed to the perturbative coupling to propagating SPhP facilitated by local surface roughness {\it via}  the excitation of near-fields with large wavenumbers. The peaks at $\lambda_{\rm II}=10.35 \mu m$ and $\lambda_{\rm III}=10.17 \mu m$ wavelengths lie within the ENZ band. Therefore, they can be ascribed to the coupling to longitudinal phonons and/or strong normal ENZ fields, although the Raman spectra did not individually resolve these features (see Supplementary Fig. 2). We remark that a high frequency resolution is needed to correctly capture these peaks {\it via}  FTIR spectroscopy, which explains why peaks $\lambda_{\rm I}$, $\lambda_{\rm II}$ and $\lambda_{\rm III}$ are not always reported in previous measurements of bare SiC substrates (see Supplementary Fig. 3 for a comparative measurement with different frequency resolutions). 

The average reflectivity for {\bf Sample II}, a processed sample with 24.0\,nm RMS roughness, is reported in Fig.\,\ref{Fig:measurementssurfaces}(f). It is found that the reflectivity of Sample II is qualitatively similar to Sample I, reflecting the same number and shape of spectral features. However, the major difference is that the resonant dip at $\lambda_{\rm I}=10.63\,\mu m$ is deepened and broadened. It can be then concluded that an increased surface roughness strengthens the coupling to propagating SPhPs, making the SPhP band particularly sensitive to surface roughness. By contrast, the reflectivity within the ENZ and dielectric bands appears to be more robust against the increase of the surface roughness.

The reduced reflectivity within the SPhP band is even more acute in {\bf Sample III}, with a RMS of 68.8\,nm (see Fig.\,\ref{Fig:measurementssurfaces}(g)). In fact, the significant reduction of the reflectivity in the SPhP band follows a strong nonlinear trend. We note that structured SiC substrates are known to support localized SPhP modes even with deeply subwavelength nanoresonators \cite{Caldwell2013}. Therefore, it should be expected that, as roughness increases, the optical infrared response of the substrate transitions from the perturbative coupling of propagating SPhPs to the excitation of localized SPhP resonances \cite{Caldwell2013}. Sample III is also characterized for the appearance of an additional spectral feature at $\lambda_{DS} = 11.92\,\mu$m, which can also be weakly appreciated in the Raman spectrum (see supplementary Fig. 2). For our 4H-SiC polytype, this wavelength is associated with the zone-folded longitudinal optical phonon (ZFLO) \cite{gubbin2019hybrid,Razdolski2016}, whose coupling to infrared radiation has been previously observed in SiC substrates structured with gratings and pillars \cite{Lu2021engineering}. The coupling of infrared far-field radiation to longitudinal optical phonons is technologically relevant as the latter can be driven with an electric bias, opening a pathway towards electrically driven polaritonic sources \cite{Lu2021engineering,Gubbin2022}. Our results demonstrate that surface roughness is capable of facilitating the coupling between SPhPs, ZFLO and far-field infrared radiation.
\newline {\bf Sample IV} presents the highest roughness, with a RMS of 298.3\,nm. The measurements reveal an even larger degradation of the reflectivity in the SPhP band, further confirming the nonlinear trend and great sensitivity against roughness of this frequency band Fig.\,\ref{Fig:measurementssurfaces}(h). As the roughness geometrical features are comparable to state-of-the-art localized SPhP resonators in SiC \cite{Dunkelberger2018,Caldwell2013}, the reduced reflectivity can be understood as the superposition of many randomly shaped resonators. It is also noticed that the dip in reflectivity in the SPhP band extends towards the ENZ band, so that the feature at $\lambda_{\rm II}=10.35\,\mu m$ is hidden. At the same time, there is a small variation of the reflectivity at the ENZ wavelength, $\lambda_{\rm LO}=10.3\,\mu m$, which now appears a peak of maximum reflectivity. Sample IV also reflects a significant reduction of the reflectivity in the dielectric bands. This effect is ascribed to the scattering by dielectric bodies of a sufficiently large size.

Overall, the reflectivity becomes a two-peaked spectrum, with one peak centred at the ENZ wavelength, and the other one centred at the TO wavelength. The first peak evidences the robustness of the ENZ band against roughness in the hundreds of nanometers scale, possibly related to the enlargement of the wavelength and inhibition of geometrical resonances. The second peak takes place at wavelengths where the real part of the permittivity of SiC is negative but very large. Thus, it hints towards the robustness of good conductors against surface roughness on the nanometer scale. We note that similar two-peaked spectra were reported for porous SiC layers \cite{Mcmillan96,RossiGiorgis}. However, these works do not provide experimental data of the transition from small to large surface roughness effects, do not discuss the coupling to different polaritonic modes, the relevance of the response around the ENZ region, and the behaviour in the dielectric bands. 

\begin{figure}
    \centering
    \includegraphics[width=0.92\linewidth]{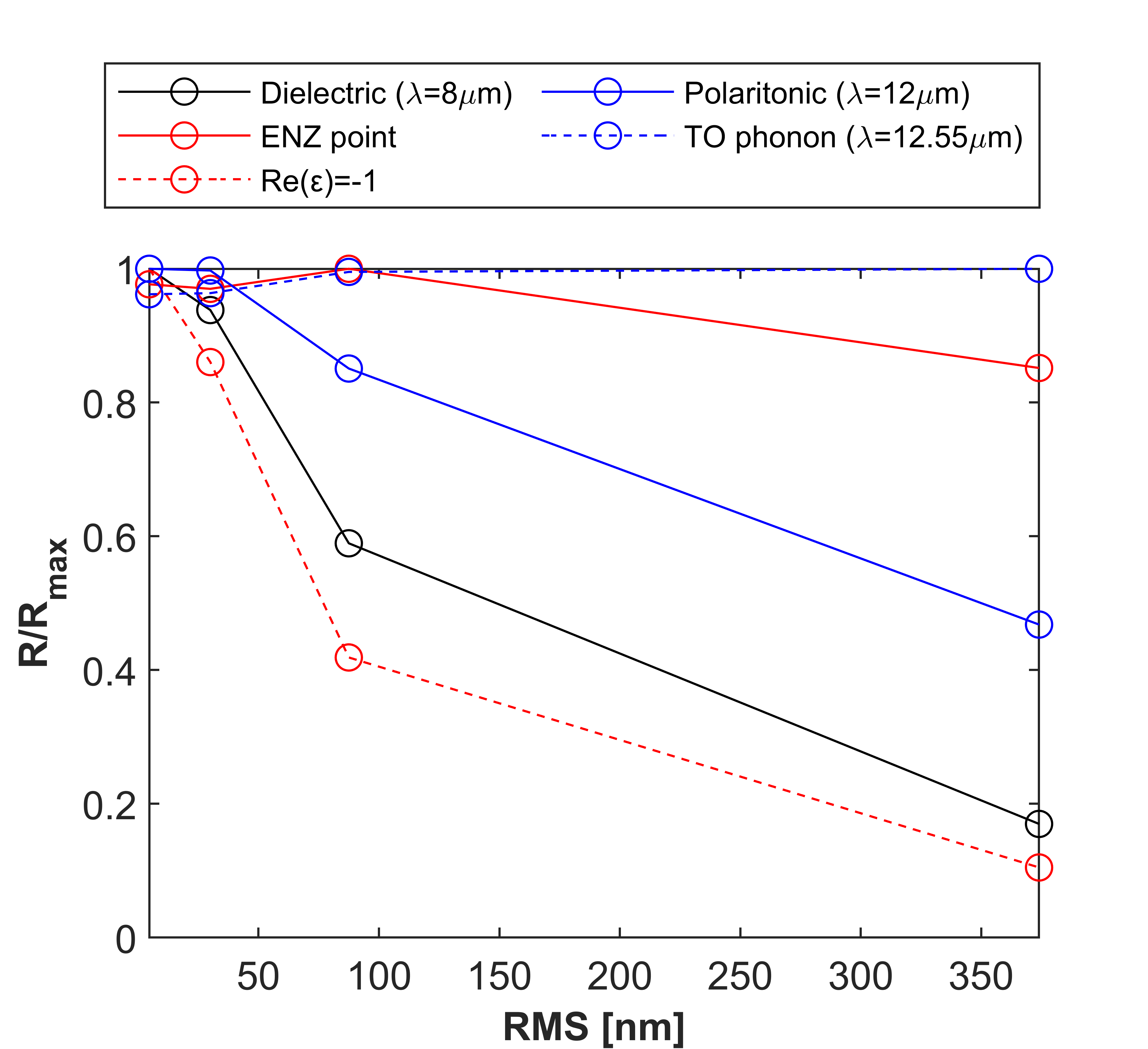}
    \caption{{\bf Normalized reflectivity values $R/R_{\rm max}$ as a function of RMS roughness for representative wavelengths}, including the dielectric band ($\lambda=8\,\mu m$, $Re(\varepsilon)= 3.83$), the ENZ point ($Re(\varepsilon)\simeq 0$, $\lambda_{\rm LO}=10.3\mu m$), the SPhP resonance ($Re(\varepsilon)\simeq -1$, $\lambda=10.56\mu m$, within the SPhP band ($\lambda=12\mu m$), and at the TO phonon point ($\lambda_{\rm TO}=12.55\mu m$).}
    \label{Fig:comparisonpoints}
\end{figure}

Fig.\,\ref{Fig:comparisonpoints} represents the measured evolution of the reflectivity as a function of RMS roughness for five representative wavelengths of $\lambda=8\,\mu m$ (dielectric band), $\lambda_{LO}=10.3\,\mu m$ (ENZ and LO wavelength), $\lambda=10.56\,\mu m$ (SPhP resonance), $\lambda=12\,\mu m$ (within the SPhP band) and $\lambda=12.55\,\mu m$ (TO wavelength). The reflectivity is normalized to its maximum value to better appreciate relative changes induced by the increased of the surface roughness. As previously discussed, the wavelengths at the SPhP resonance and within the SPhP band present the fastest decrease of the reflectivity, where now the nonlinear trend with the RMS can be clearly appreciated. The normalized reflectivity in the dielectric band is also found to exhibit a strong decrease with surface roughness, an effect that could not be easily identified in the measured spectra due to its low initial value. By contrast, the ENZ (LO) wavelength and TO wavelengths are shown to be particularly robust to the increase of the RMS roughness from few to hundreds of nanometers.

Our experiments clarify the impact of surface roughness on ENZ substrates, and its comparison with other material responses. It is found that, for small surface roughnesses, ENZ is sensitive to surface roughness as it facilitates the coupling to longitudinal phonons and/or strong normal fields in the ENZ band, as suggested by previous theoretical reports \cite{Javani2016real}. At the same time, such coupling does not seem to significantly decrease the reflectivity and it does not severely increase along with the RMS roughness. In fact, for RMS roughnesses in the range of tenths and hundreds of nanometers the ENZ band appears to be particularly robust, as compared to the SPhP and dielectric bands, in connection with the enlargement of the wavelength and the geometry-invariant phenomena observed in ENZ media \cite{Liberal2016geometry}.

\begin{figure}
    \centering
    \includegraphics[width=0.99\linewidth]{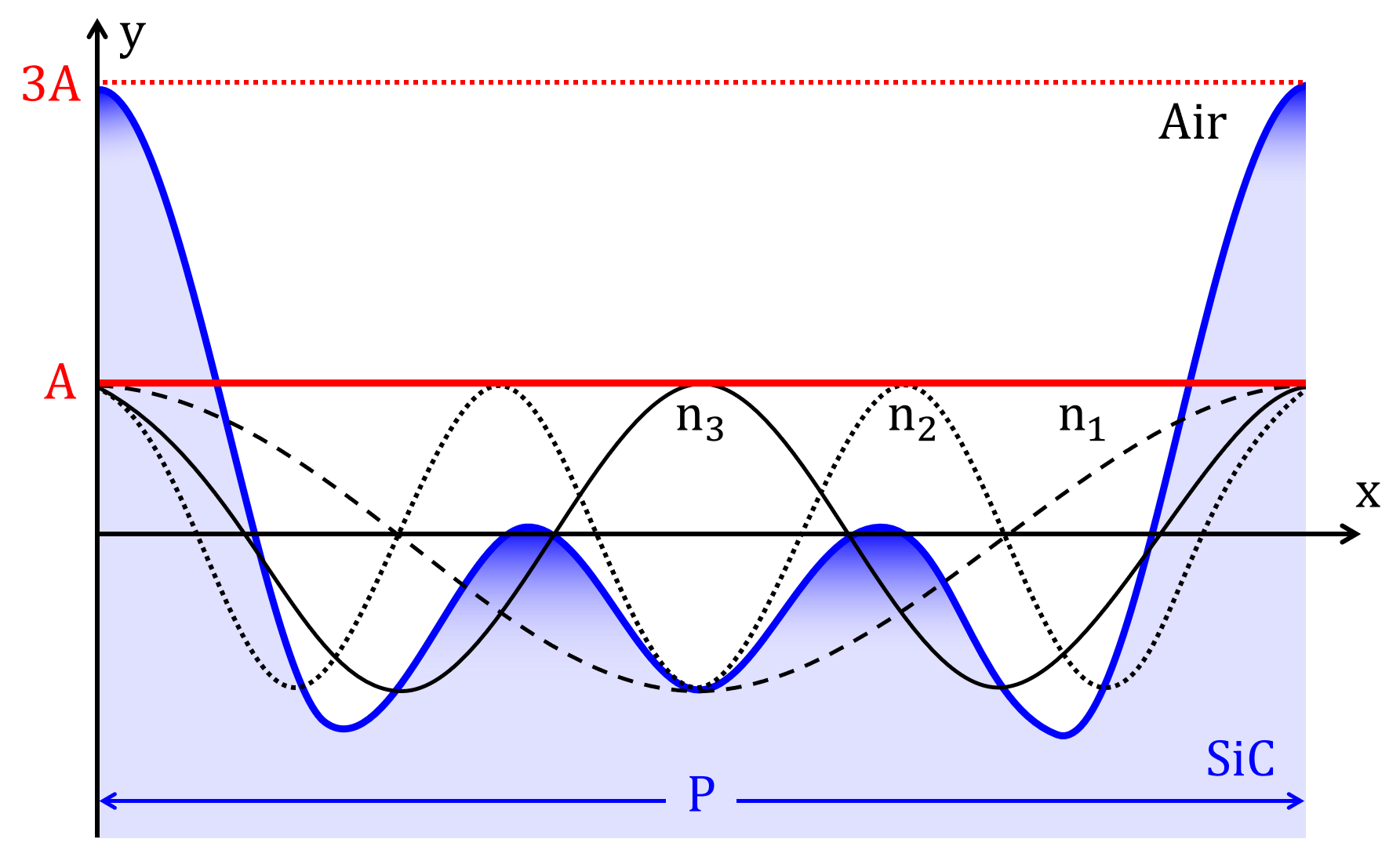}
    \caption{{\bf Simplified geometrical model for surface roughness.} The geometry of the rough surfaces with mimicked in a unit cell of length $P$ with the superposition of three sinusoidal functions with the same amplitude $A$, and spatial frequencies $n_1=1/P$, $n_2=2/P$ and $n_3=3/P$.}
    \label{Fig:simulatedwave}
\end{figure}

\section{Numerical Simulations}

We use full-wave numerical simulations to provide additional insight on the impact of surface roughness in SiC substrates. The theoretical modelling of surface roughness is a challenging task. On the one hand, the interaction of infrared radiation results in complex electromagnetic excitations and resonant effects, which cannot always be captured by effective medium theories (EMT) \cite{Bruggeman,Monecke_1994,MaxwellGarnett,LOOYENGA1965401}. On the other hand, numerical simulations can fully capture the wave effects introduced by the geometry, but the computational burden required to reproduce the geometrical details of random roughness can be too demanding. In the following, we use full-wave numerical simulations with a simplified geometry (see Methods), a technique that has been successfully employed to address the impact of surface roughness in plasmonic cavities \cite{Ciraci2020impact,Macias2016surface}.

The simulation setup is schematically depicted in Fig.\,\ref{Fig:simulatedwave}. It consists of a periodic two-dimensional (2D) model, where the surface roughness is built from the sum of sinusoidal functions according to the following equation: $y_R(x)=\sum_{n=1}^{3}A \,cos(2\pi\,x\,nP^{-1})$. $A$ is the amplitude of the sinusoidal functions, $x$ the displacement along the $X$ axis, and $P$ the length or pitch of the periodic unit cell. The configuration is excited with a normally incident plane wave with transversal magnetic (TM) polarization, i.e., with an in-plane electric field, as such polarization provides the stronger interaction with nanometer changes in the geometry of the surface. Our 2D periodic model is clearly an approximation, but it qualitatively captures the main physics of the configuration while substantially reducing the computational burden. In turn, the reduction of the computational burden allows for extensive parameter sweeps mimicking the stochastic nature of surface roughness. Specifically, we estimate the electromagnetic response of surface roughness by sweeping the period of the unit cell from 10\,nm to 900\,nm, with a step of 20\,nm, and then averaging the resulting reflectivities. Within our model, SiC is modelled with an isotropic permittivity following a Lorentzian dispersion profile (see Methods). Therefore, the numerical simulations cannot predict responses not contained within such material model, e.g., the coupling to ZFLO phonons.

\begin{figure}
\centering
\begin{overpic}[width=0.95\linewidth]{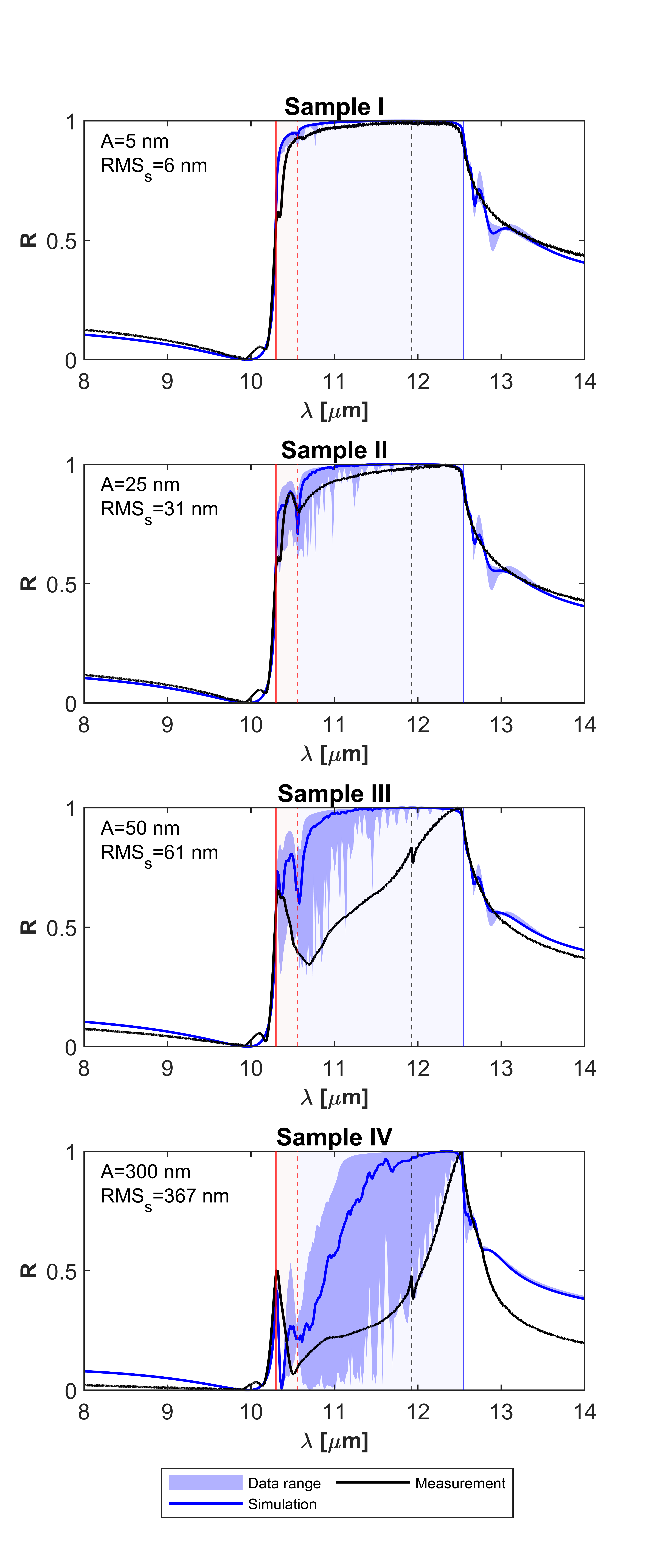}
    \put(6, 28){\small{(d)}}
    \put(6, 50){\small{(c)}}
    \put(6, 72){\small{(b)}}
    \put(6, 94){\small{(a)}}
    \end{overpic}
    \caption{{\bf Numerical results.} Comparison between the reflectivity predicted by the numerical simulations (blue line) and the measured reflectivity spectra (black line). The data range expanded by all numerical simulations is shown as a blue band between the minimum and maximum values recorded on each individual simulations.}
    \label{Fig:simulationresults}
\end{figure}

The predicted reflectivities are gathered together in Fig.\,\ref{Fig:simulationresults}, including the average reflectivities for unit-cell periods ranging from 10\,nm to 900\,nm with a 20\,nm step, and the data range expanded by all individual simulations (shown as a blue band between the minimum and maximum values). The reflectivity data for all individual simulations is reported in Supplementary Figs.\,4 and 5. We report the results for four different amplitudes A=5\,nm, 25\,nm, 50\,nm and 300\,nm, corresponding to RMSs of 6.1\,nm, 30.6\,nm, 61.2\,nm and 289\,nm, on the same range as the fabricated samples. Overall, it is found the numerical simulations qualitatively agree with the experiments. The simplifications of the numerical model (2D vs 3D geometry and use of a periodic and subwavelength unit-cell) do not allow for a quantitative matching between theory and experiments. However, the predictions from the numerical model qualitatively reproduce the measured spectra (existence and positioning of the peaks), thus supporting the main conclusions drawn in the previous section. For example, the predicted reflectivities for the two first samples with amplitudes of $A=$ 5\,nm and 25\,nm (see Fig.\,\ref{Fig:simulationresults} (a)) are predominantly characterized for a dip near the SPhP resonance frequency, where $Re(\varepsilon_{\rm SiC})=-1$, confirming the sensitivity of the SPhP resonance frequency against surface roughness. On the other hand, the numerical simulations do not predict the dips experimentally observed in the ENZ band. However, a small ripple in the reflectivity can be appreciated at a wavelength of $\lambda_{\rm II}=10.35\,\mu m$, consistent with one of the measured dips in the ENZ band.

\begin{figure*}
\centering
\begin{subfigure}{0.31\linewidth}
    \begin{overpic}[width=\linewidth]{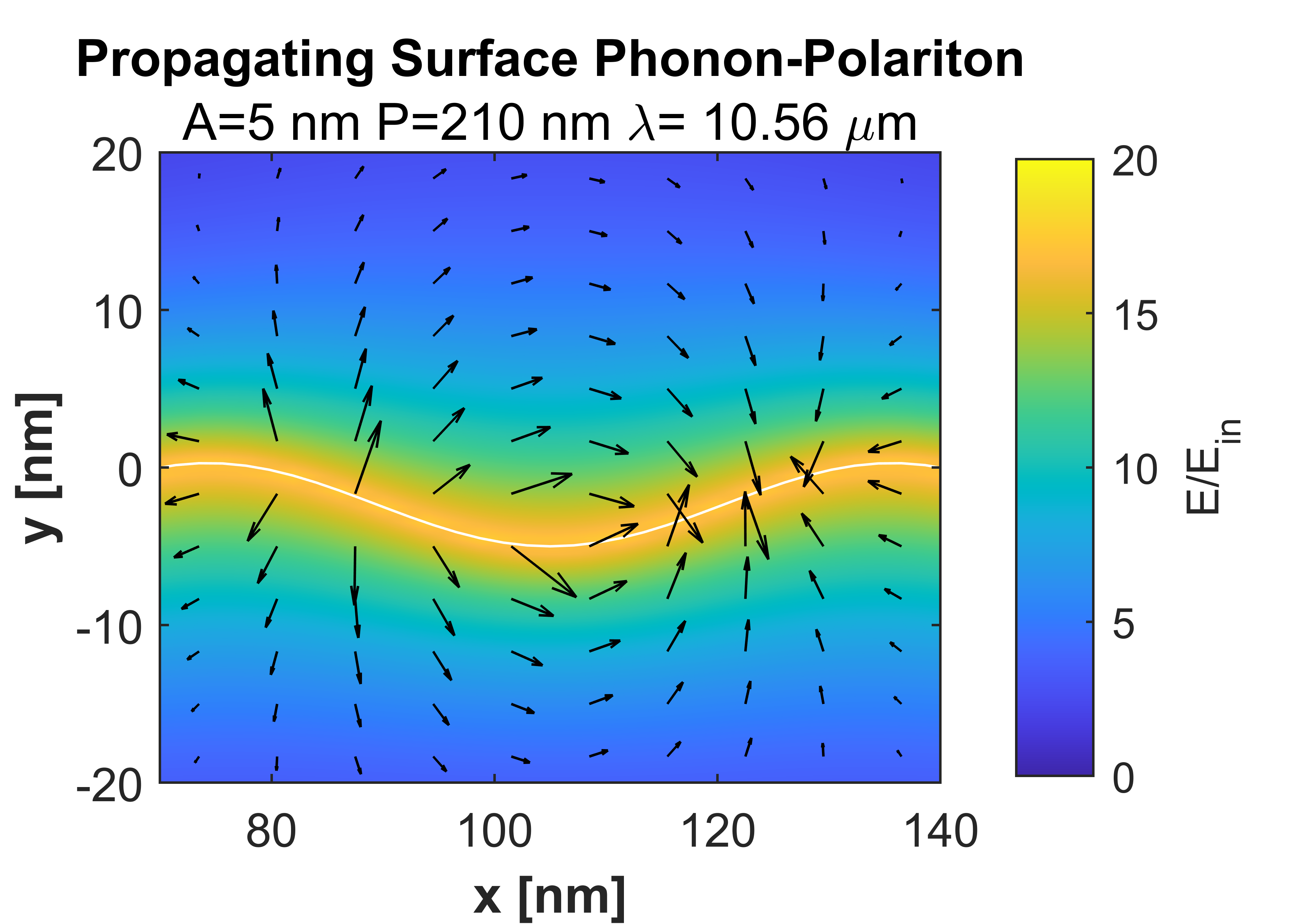}
    \put(-6, 60){\small{(a)}}
    \end{overpic}
    \end{subfigure}
    \begin{subfigure}{0.31\linewidth}
        \begin{overpic}[width=\linewidth]{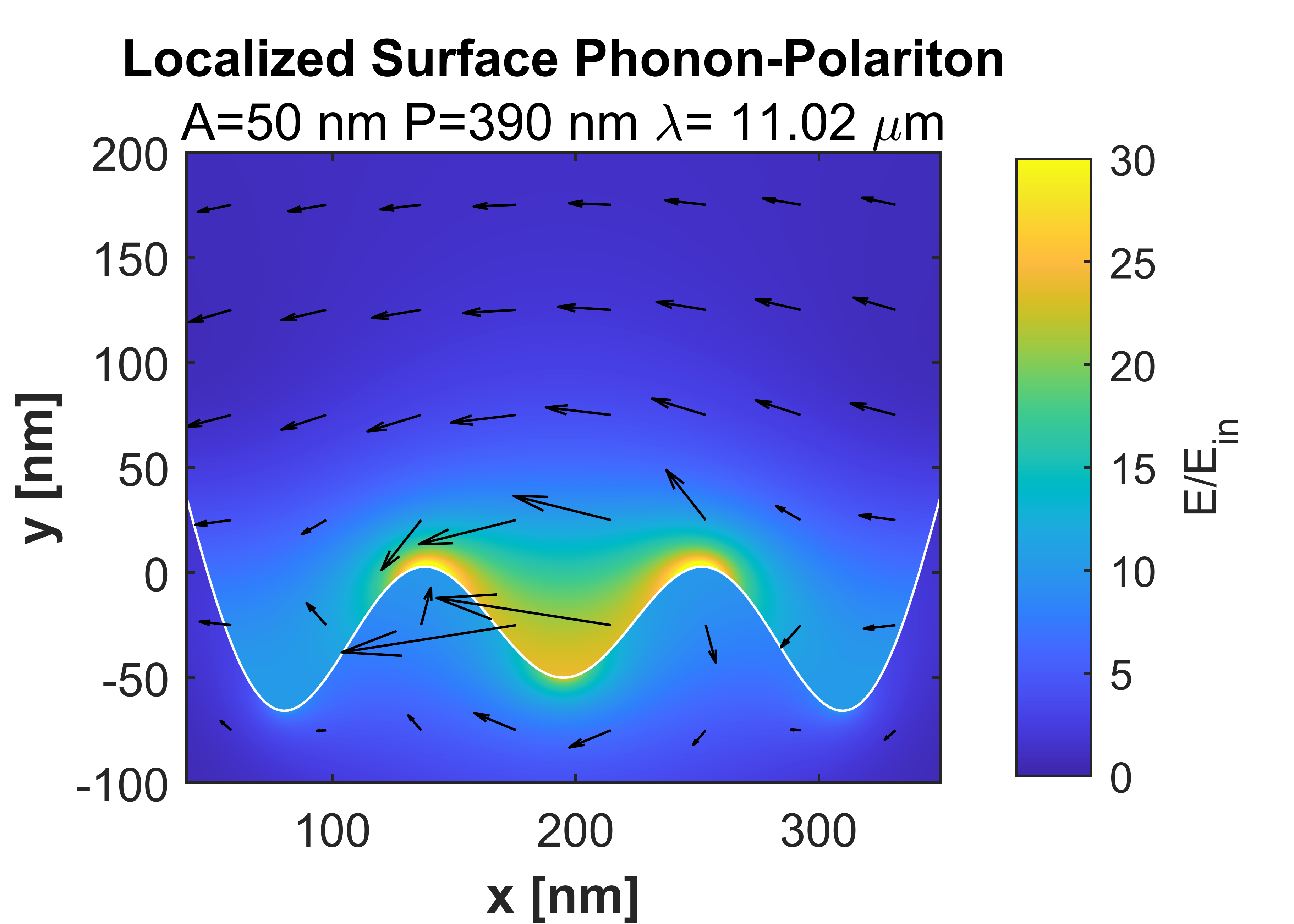}
    \put(-6, 60){\small{(b)}}
    \end{overpic}
    \end{subfigure}
    \begin{subfigure}{0.31\linewidth}
        \begin{overpic}[width=\linewidth]{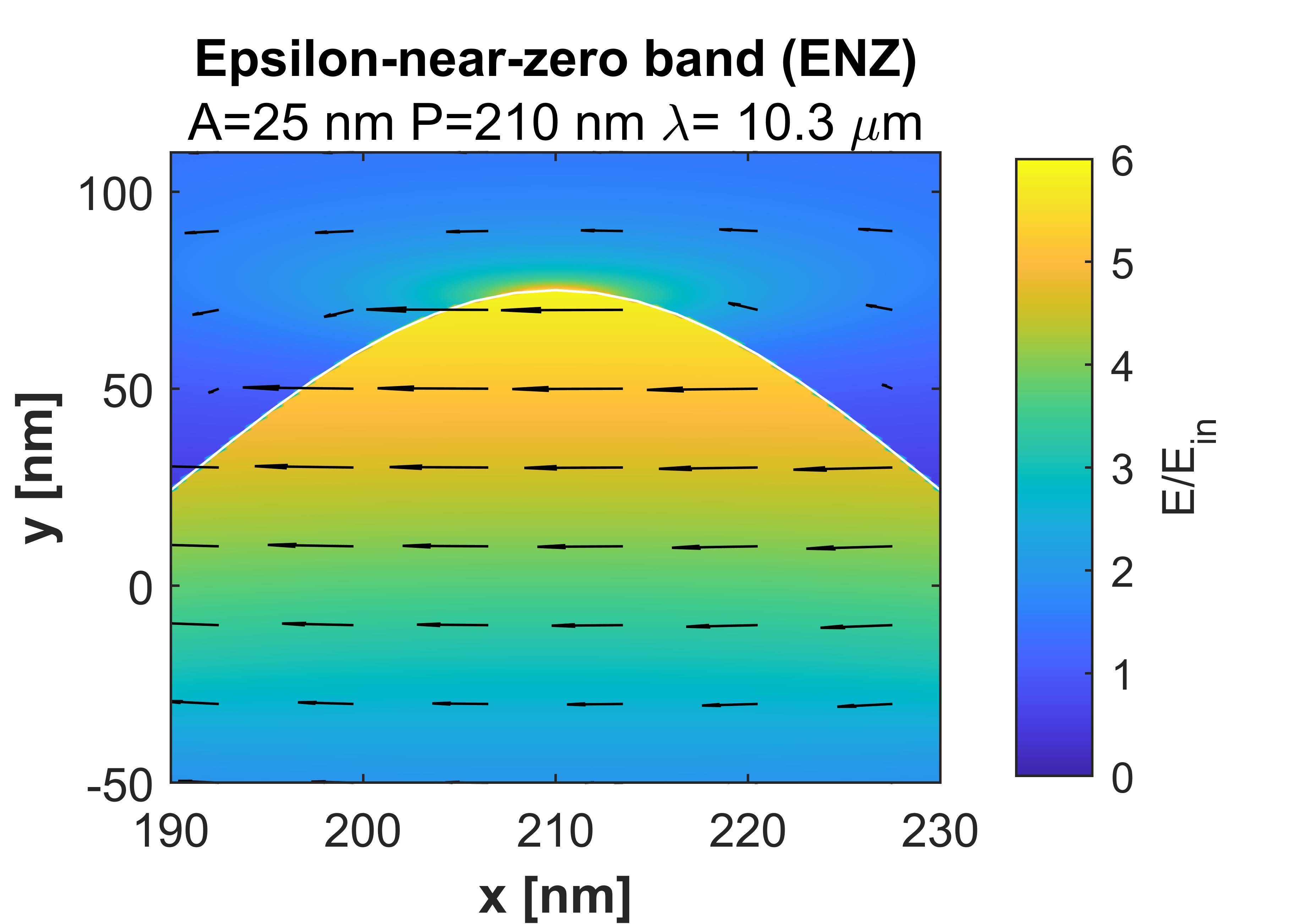}
    \put(-6, 60){\small{(c)}}
    \end{overpic}
    \end{subfigure}
    \begin{subfigure}{0.31\linewidth}
        \begin{overpic}[width=\linewidth]{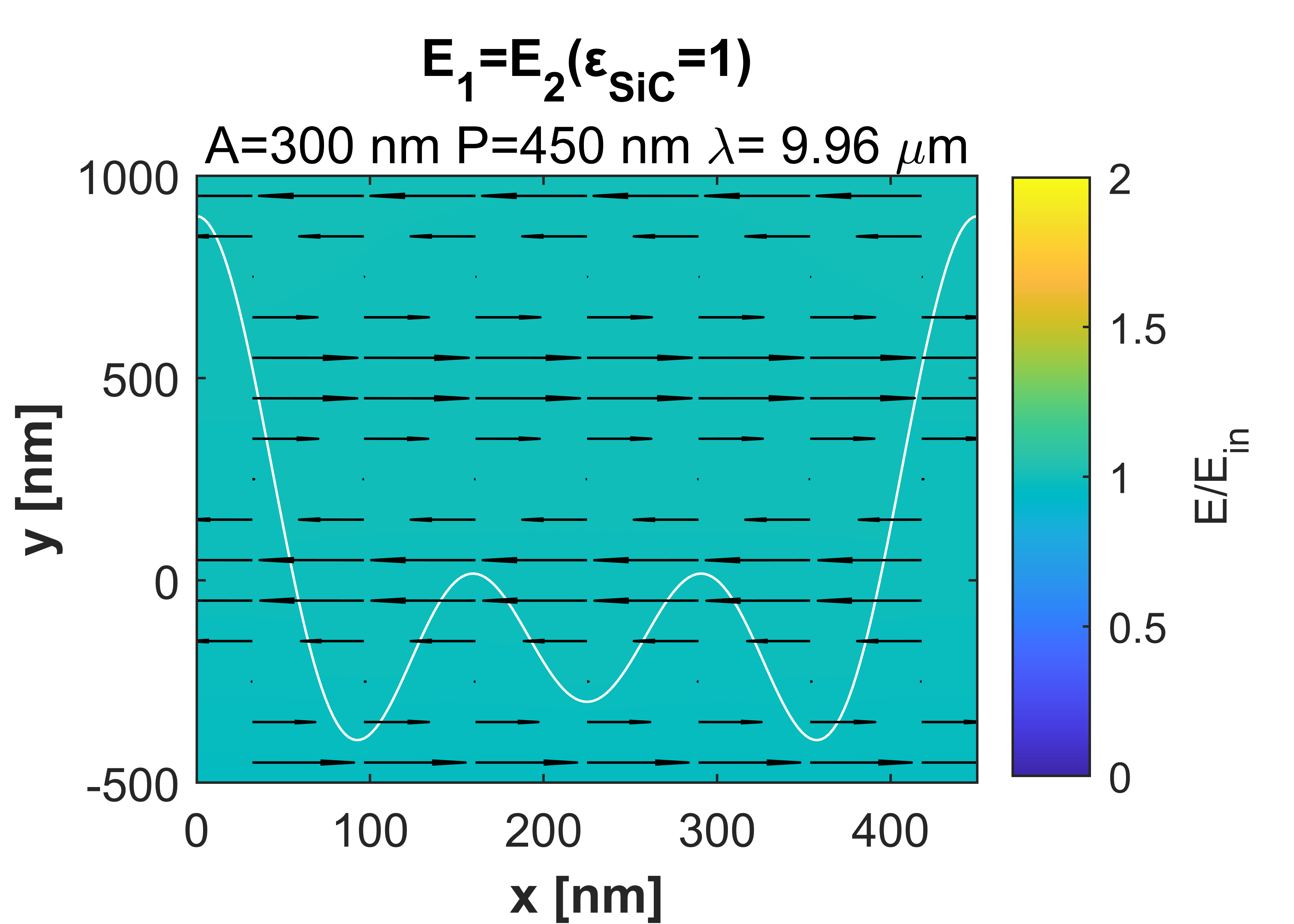}
    \put(-6, 60){\small{(d)}}
    \end{overpic}
    \end{subfigure}
    \begin{subfigure}{0.31\linewidth}
        \begin{overpic}[width=\linewidth]{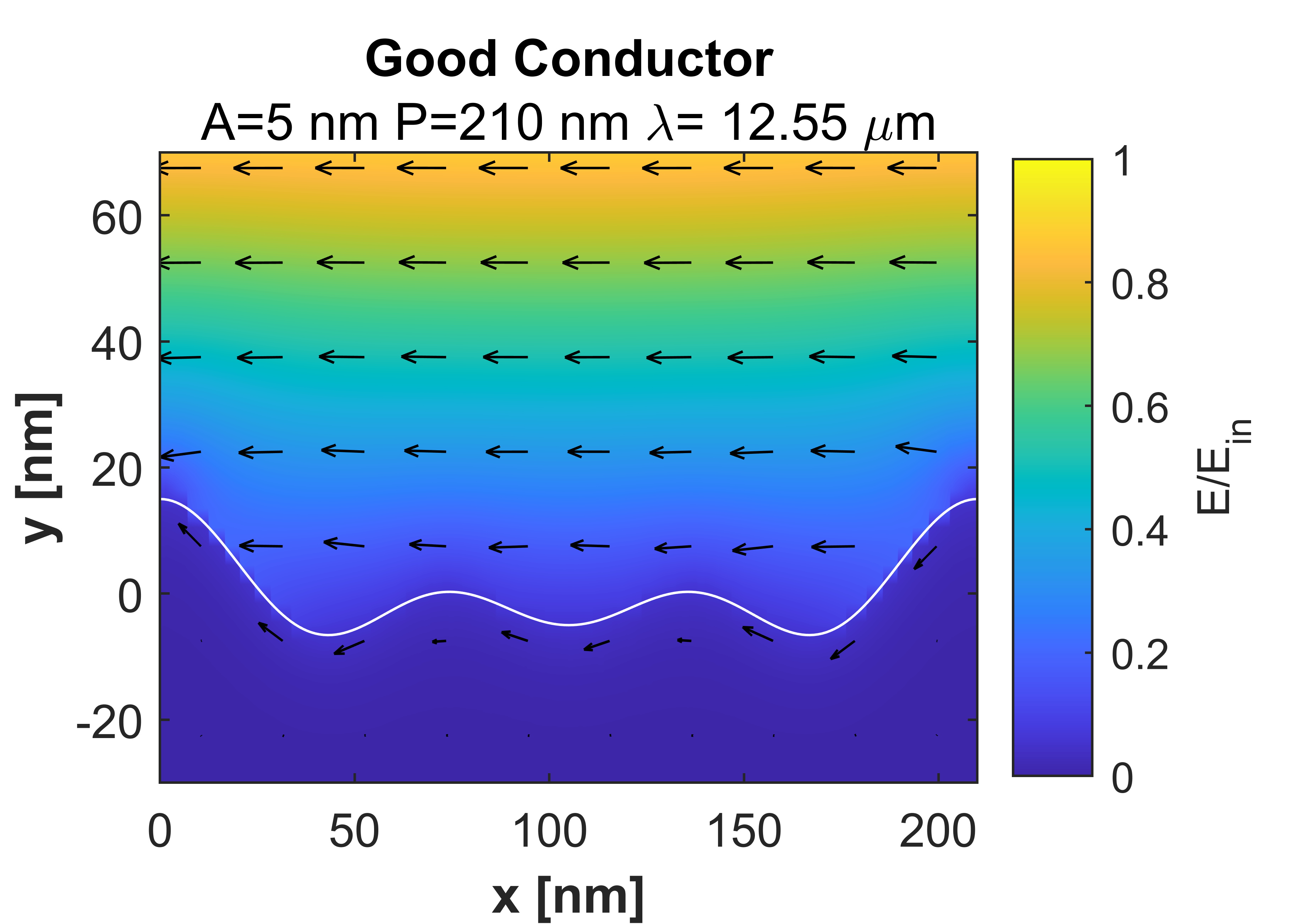}
    \put(-6, 60){\small{(e)}}
    \end{overpic}
    \end{subfigure}
      \caption{{\bf Examples of computed field distributions for the polaritonic phenomena excited at a rough a silicon carbide (SiC) substrate}. (a)-(e) Magnitude (colormap) and vectorial field (arrows) of the electric field distribution for selected examples corresponding to: (a) coupling to propagating SPhPs, (b) the excitation of localized SPhPs, (c) strong longitudinal fields and the ENZ wavelength, (d) suppression of the scattering at $\varepsilon_{SiC}\simeq 1$, and (e) approximate good conductor response at the TO wavelength. The wavelength $\lambda$, amplitude $A$ and unit-cell length $P$ values for each example are indicated on top of the figures.}
    \label{fig:localizedsurfacephononpolaritons}
    \end{figure*}
    
The dip in the SPhP band deepens and extends its frequency range for the numerical simulations with an amplitude of $A=50$\,nm, qualitatively following the measurements of Sample III. At the same time, the numerical simulations are found to quantitatively underestimate the impact of the surface roughness in the SPhP band. Numerical simulations for this sample correctly recover the dip in the ENZ band measured at $\lambda_{\rm II}=10.35\,\mu m$, suggesting that the measured dip could be related to the optical response in the ENZ band of SiC. However, the dip at $\lambda_{\rm III}=10.17\,\mu m$ is not captured by the numerical simulations. Thus, its origin can be ascribed to the coupling to additional longitudinal phonons and/or as a signature of anisotropy in the sample, both effects not included in our isotropic permittivity model (see Methods). Similarly, the numerical simulations are not capable of reproducing the coupling to the ZFLO phonon at $\lambda = 11.92\,\mu$m. Finally, the numerical simulations for an amplitude of $A=$300\,nm correctly predict a two-peaked spectrum as experimentally observed for Sample IV. The numerical simulations further confirm that the peaks of maximal reflectivity are exactly centered at the LO (ENZ) and TO phonon wavelengths, supporting the robustness of both frequency points towards roughness effects in the tenths and hundreds of nanometers scale.
    
Similar to the experiments, the data range expanded by the numerical simulations increases along with the surface roughness and within the SPhP band. In the case of the simulations, the effect is stronger and it has a clear physical origin. Resonant effects associated with localized SPhP resonances appear for a sufficiently large roughness, which take place in discrete points that shift its wavelength for each simulated length of the unit cell $P$ (see also Supplementary Fig. 4). These spectral features disappear when averaging the reflectivies for many $P$ values, but their presence is evidenced by an extended data range. Therefore, the extended data range supports the prevalence of localized resonance for substrates with a large roughness, and the interpretation of the broadened dip in the SPhP band as a continuum distribution of subwavelength resonators.

Another advantage of full-wave numerical simulations is that they give access to the field distributions excited in the structure, providing additional physical insight. Fig.\,\ref{fig:localizedsurfacephononpolaritons} represents the electric field distributions for a few selected examples that confirm the nature of the polaritonic phenomena discussed in previous sections and schematically depicted in Fig.\ref{Fig:ConceptIntro}. The reflectivities predicted for these examples are reported in Supplementary Fig.\,4.
For example, at the SPhP resonance ($Re(\varepsilon_{SiC})=-1$), the predicted field distribution confirms the excitation of propagating SPhP resonances for small roughnesses. Specifically, the field is concentrated along the entire surface, evidencing the perturbative excitation of a propagating SPhP at the interface (see Fig. \ref{fig:localizedsurfacephononpolaritons}(a)). By contrast, the propagation of the SPhP along the interface is interrupted as the surface roughness amplitude increases. This effect leads to the formation of localized SPhP resonances, as it is most clearly appreciated in the configuration shown in Fig. \ref{fig:localizedsurfacephononpolaritons}(b), where the roughness geometry conforms a nanoresonator. As shown in Supplementary Fig.\,4, the excitation of this localized SPhP is associated with a narrowband dip in the prediction of the reflectivity.

At the ENZ wavelength, the electric field distribution is confirmed to be dominated by strong electric fields within the SiC substrate (see Fig.\,\ref{fig:localizedsurfacephononpolaritons}(c)). As anticipated, this effect arises from the fact that the  normal displacement vector must be continuous across the interface, $\mathbf{D}_{air,n}=\mathbf{D}_{SiC,n}$, such that $\varepsilon_{air} \mathbf{E}_{air,n}=\varepsilon_{SiC} \mathbf{E}_{SiC,n}$. Therefore, when the permittivity of SiC approaches zero $\varepsilon_{SiC}\simeq 0$, the electric field on the SiC side of the interface must be much larger than on the air side $\mathbf{E}_{SiC,n}\gg\mathbf{E}_{air,n}$. Another interesting example takes place when $\lambda=9.96\,\mu m$, and the permittivity of SiC approaches unity ($Re(\varepsilon_{SiC})\simeq 1$, as shown in Fig. \ref{fig:localizedsurfacephononpolaritons}(d)). In such a case, all scattering effects at the surface are suppressed, and the incident wave is smoothly transferred into the substrate, justifying the dip in reflectivity observed in all experimental samples at this wavelength. Finally, at the TO wavelength ($\lambda=12.55\,\mu m$, Fig.\,\ref{fig:localizedsurfacephononpolaritons}(e)), the large value of the permittivity repels the fields outside the SiC substrate, so that SiC behaves similarly to a good conductor. In fact, for small amplitudes of the surface roughness, the electric field if predominantly transversal and a local minimum appears close to the SiC interface, further confirming a response analogous to a good conductor. 
    
\section{Conclusion}

Our results provide a comparative view of the impact of surface roughness on the reflectivity of SiC substrates across different size-scales. It was found that the interplay between the complex roughness geometry and the dispersive material response of SiC gives rise to a variety of polaritonic effects. These include propagating and localized SPhP resonances, strong normal fields emerging from ENZ boundary conditions, as well as the coupling to zone-folded and longitudinal optical phonons. The experimental results were theoretically supported with full-wave numerical simulations with a simplified geometry and extended parameter sweeps. The results provided by the numerical simulations qualitatively match the measurements, providing further confirmation of the conclusions drawn from reflectivity measurements, and provide additional physical insight into the polaritonic phenomena associated with the interaction of infrared radiation with a rough SiC substrate. 

The dispersive material response of SiC  contains dielectric, SPhP and ENZ bands with the same sample, which allowed us to compare the sensitivity of different material responses in a controlled experiment. It was found that the ENZ band is negatively affected by the coupling to longitudinal phonons and strong normal electric fields in the presence of nanometric surface roughness. At the same time, it was concluded that for roughnesses with size-scales in the hundreds of nanometers, the ENZ band is more robust than the dielectric and SPhP bands. We believe that these results provide an important step forward in understanding the robustness and limitations of realistic ENZ materials for nanophotonic technologies. 
In addition, our results demonstrate that artificially-induced surface roughness enables the coupling between ZFLO phonon, SPhP and farfield radiation, which might be useful for the design of electrically driven sources without the need of fabricating nanoresonators \cite{Lu2021engineering,Lu2021engineering,Gubbin2022}. Moreover, it was shown that engineering artificially-induced roughness controls the coupling to both propagating and localized SPhP polaritons, pointing towards the design of litography-free thermal emitters \cite{Ghobadi2022lithography}.

\section{Methods}

\subsection{Theory and numerical Simulations}

 The calculations had been carried out with the \textit{frequency domain solver} of the \textit{Wave Optics module} of the full-wave solver Comsol Multiphysics \texttrademark \cite{COMSOL}. The roughness then is defined as sum of three waves in order to simulate a periodic structure (\textit{Periodic Boundary Conditions}). The bottom limit is set as an \textit{Scattering Boundary Condition}. Silicon Carbide was modelled with a permittivity following the Lorentzian dispersion profile reported in Fig.\,\ref{Fig:evanescentpropagating}(b), with $\varepsilon_{SiC}(\omega) = \varepsilon_{\infty}(\omega^2 - \omega_p^2 + i\omega\omega_c)/(\omega^2 - \omega_0^2 + i\omega\omega_c)$, with fitted values $\omega_p = 3.93\,$x$\,10^{15}$\,rad/s, $\omega_0 = 1.5\,$x$\,10^{15}$\,rad/s and $\omega_c = 1.06\,$x$\,10^{15}$ rad/s \cite{Caldwell2015}. 

In Fig.\,\ref{Fig:evanescentpropagating}(c), the SPhP dispersion profile was calculated following the equation: $k_{SPhP}=(\frac{\omega}{c})\,\sqrt{\frac{\varepsilon_{SiC}\,\varepsilon_{air}}{\varepsilon_{SiC}+\varepsilon_{air}}}$, while bulk propagating modes followed $k=\omega/c\sqrt{\varepsilon_{SiC}}$ \cite{Caldwell2015}.

\subsection{Samples fabrication and measurement}

The samples were fabricated in the ISO7 cleanroom at UPNA facilities, {\it via} deep reactive ion etching (DRIE). The normal incidence reflection was measured with a microscope Hyperion 3000  under Fourier Transform Infrared Spectroscopy (FTIR).\\
Surface roughness characterization is achieved with an atomic force microscope (AFM) images in areas of $20\,\mu m\,\times\,20\,\mu m$ of 256 lines, and the later data subjected to a polynomial correction due to a natural mismatch created by the microscope.    

\bibliography{library}

\end{document}